\documentclass[12pt,preprint]{aastex}

\shorttitle{SDSS-XMM Quasars}
\shortauthors{Risaliti \& Elvis}

\begin{document}

\title{The SDSS-XMM Quasar Sample: First Results} 

\author{Guido Risaliti\altaffilmark{1,2} \&  Martin Elvis\altaffilmark{1}}
\email{grisaliti@cfa.harvard.edu}

\altaffiltext{1}{Harvard-Smithsonian Center for Astrophysics, 60 Garden St. 
Cambridge, MA 02138 USA}
\altaffiltext{2}{INAF - Osservatorio Astrofisico di Arcetri, L.go E. Fermi 5,
Firenze 05125 Italy}
%%%%%%%%%%%%%%%%%%%%%%%%%%%%%%%%%%%%%%%%%%%%%%%%%%%%%%%%%%%%%%%%%%%%%%%
\begin{abstract}
We have searched in the {\it XMM-Newton} public archive for quasars in
the Sloan Digital Sky Survey (SDSS) First Data Release (DR1), and
found 55 lying in the field of a {\it XMM-Newton} observation with
exposure times $>$20~ksec (as of August, 2004). The 35 quasars which
yielded good X-ray spectra span redshifts from 0.5 to 2.5.  The large
collecting area of {\em XMM-Newton} allows us to investigate the
dependence of the X-ray spectra of quasars on luminosity, redshift and
optical colors. We find: (1) no evolution of X-ray slope ($\Gamma$)
with either redshift or luminosity; (2) no correlation between $\Gamma$ or
absorbing column density and the optical to X-ray ratio,
$\alpha_{OX}$; (4) no relation between $\alpha_{OX}$ and optical
colors.  The two latter results suggest that obscuration is not the
dominant cause of the spread in X-ray slope or optical color. 
We find four unusual quasars, 10\% of the sample: three are absorbed
(N$_H>$10$^{22}$cm$^{-2}$), of which one has high luminosity
(1.5$\times$10$^{44}$erg~s$^{-1}$); the fourth has
$\Gamma$=0.6$\pm$0.2, far flatter than the typical value of 1.8-2.0,
and a strong emission line (EW=1.2$\pm$0.4~keV) which, if Fe-K, implies
a redshift of $\sim$1.4.
\end{abstract}


%%%%%%%%%%%%%%%%%%%%%%%%%%%%%%%%%%%%%%%%%%%%%%%%%%%%%%%%%%%%%%%%%
\section{Introduction}

Surveys of the X-ray properties of optically selected quasars have
followed two main lines: 
\newline\noindent 
(1) detailed spectral analysis of relatively small
samples ($\leq$30) of bright, mostly low redshift objects (Elvis et
al. 1987, Laor et al.~1997, George et al. 2000, Reeves \& Turner
2000), primarily using the Bright Palomar Survey ($B<16.3$) (``PG'',
Schmidt \& Green 1983).  These works have shown that radio quiet
quasars have on average a power-law spectrum in the X-rays, with
photon index $\Gamma\sim2$, analogous to what is found for the lower
luminosity Seyfert 1 Galaxies (George et al. 1998, Perola et
al. 2002), and are not absorbed, with the exception of Broad
Absorption Line (BAL) quasars, which represent $\sim10-20$\% of
optically selected objects (Reichard et al. 2003).
\newline\noindent 
(2) statistical investigations of large ($\sim100$) samples using only
X-ray fluxes, in order to compare X-ray and optical/UV properties
(Zamorani et al. 1981, Avni \& Tananbaum 1986, Wilkes et al. 1994,
Yuan et al. 1998, Bechtold et al. 2003, Vignali, Brandt \& Schneider
2003, hereinafter V03, Strateva et al.~2005). 
These works have clearly established a
relation between the X-ray to optical ratio with optical luminosity:
more luminous objects are X-ray fainter (relatively to the optical
emission).
\newline\noindent 
%
%Up to now these two approaches have had strong limitations: the first
%approach has not been possible for large or well selected samples of
%high redshift, high luminosity objects due to the lack of good quality
%spectra, while the second approach made use only of X-ray fluxes (or
%upper limits) and no spectral information was available (except, in a
%few cases, for hardness ratios).  

What has not been possible is a study of the X-ray spectra of a large
sample of quasars, especially one spanning a wide range of luminosity
and redshift.
Hence a change in X-ray loudness
might be due to a change of the intrinsic continuum production method
or efficiency, or merely to line of sight obscuration (Risaliti et
al.~2001).

The new generation optical surveys and X-ray instruments are changing
this situation significantly: the Sloan Digital Sky Survey (SDSS, York
et al.~2000) will provide high quality optical spectra and photometry
for $\sim100,000$ quasars spanning a redshift interval from 0 to $>6$.

The expected X-ray flux of SDSS quasars can be estimated from optical
magnitudes assuming the median standard Spectral Energy Distribution
(SED) of Elvis et al. (1994) and the luminosity - $\alpha_{OX}$
relation of V03. They range from $\sim10^{-14}$ to
$\sim10^{-13}$~erg~s$^{-1}$~cm$^{-2}$, accessible to {\em Chandra}
(Weisskopf et al.~2002) and {\it XMM-Newton} (Jansen~et~al.~2001) for
good quality X-ray spectra ( $>$100-1000 counts).  Pointed
observations with both the X-ray observatories have been performed for
the highest redshift ($z>4$) SDSS quasars (Brandt et al. 2002,
Bechtold et al. 2003, Vignali et al. 2003a,b).

To obtain large samples of X-ray spectra for lower redshift SDSS
quasars targetted observations are less efficient than a
`serendipitous' approach.  {\it XMM-Newton} has larger collecting area
($\sim1,800$~cm$^2$ at 2~keV, vs. $\sim400$~cm$^2$) and field of view
(a circle of $\sim$14 arcmin radius, vs $\sim$8~arcmin) than {\em
Chandra}. 
%(Vignetting is not severe: at 12~arcmin from the on-axis
%position the effective area decreases by a factor of $\sim2$.)  
A source with a standard radio-quiet quasar spectrum and 2-10~keV flux
of 10$^{-14}$~erg~s$^{-1}$~cm$^{-2}$ observed by XMM will give a
$\sim100$ counts in $\sim$10~ksec.

Here we present the results of a first analysis of the X-ray properties
of 55 SDSS quasars serendipitously observed by {\em XMM-Newton} in long
($>20$~ksec) observations. The quasars span redshifts from 0.5-2.5 and a
wide range of luminosities.
The SDSS selection criteria (Richards et al. 2002) are based on the
locus of quasars in a multi-color space, and allows the selection of
objects redder than the classical UV-excess selected objects
(e.g. Schmidt \& Green 1983). As a consequence, a significant fraction
($\sim10$\%) of optically red quasars is present in the SDSS (Richards
et al. 2003).
We first investigate the relations between X-ray properties and luminosity
\footnotetext{We estimate luminosity distances using the standard
``concordance'' cosmological parameter $(h_0, \Omega_M,
\Omega_\lambda)=(0.7,0.3,0.7)$, Spergel et al.~2003.}, 
redshift and optical colors. We then concentrate on the analysis of four objects
with peculiar X-ray properties. Finallly, we briefly discuss the expected
quality of the final sample.
% and so, with the more general color
%selection methods of SDSS, they allow an analysis of the relations
%between X-ray spectral properties of quasars and redshift, luminosity,
%optical colors (Section 3).  We then present a detailed estimate of
%the quality of the final SDSS/XMM sample (Section 3.2).
%

%%%%%%%%%%%%%%%%%%%%%%%%%%%%%%%%%%%%%%%%%%%%%%%%%%%%%%%%%%%%%%%%%%%%%%%%%%%%%%
\section{The SDSS - XMM-Newton DR1-04 Quasar Sample}

As of August 2004, of the 18,650 quasars of the SDSS DR1, the {\em
XMM-Newton} public archive\footnote{ URL: http://xmm.vilspa.esa.es} 
contained 150 within 14' of the XMM
optical axis, which are not the main targets of the XMM observations.
% ensuring that none were the targets of the {\it
% XMM-Newton} observations. (I.e. {\it all} the sources are
% serendipitous).  
The length of the XMM observations varies between
$\sim$4~ksec and $\sim$100~ksec.
%
%For each quasar in this `DR1-04' sample we estimated the expected
%detectability of the object using the X-ray flux predicted from the
%optical U and B magnitudes, using the mean VO3 relation between
%$\alpha_{OX}$ and the optical luminosity\footnotemark,
%
%
%We then estimated the expected net number counts and S/N 
%and taking into account the mirror vignetting and the average
%background counts.  Some $\sim$80\% of the SDSS quasars should be
%detected, and in for $\sim60$\% the S/N should be enough ($>10$) to
%allow a spectral analysis.

%The distribution of $\alpha_{OX}$ in the SDSS/XMM `DR1/04' sample is
%consistent with that of Yuan et al.~1998. However, such 

We extracted the XMM data for the 58 quasars in the 20
longest {\em XMM-Newton} observations ($>20$~ksec).  For each object
we extracted the spectra from the two EPIC CCD detectors PN and MOS (from
the combined MOS1 and MOS2 observations). In 15 cases the
object was found to lie in a bad region in one of the two instruments,
either in a strip between two adjacent chips, or in a zone outside the
field of view (in particular, the PN instrument has a rectangular
shape, therefore our circular selection region does not overlap at
large distances [$>10'$] from the on-axis position). For these 15
objects we performed our analysis using the only available detector.
In 3 cases ($\sim5$\%) the quasar lay a bad region in both the
observations, therefore no XMM spectrum is available.  For the
remaining 40 cases both the PN and MOS spectra are available.  For
five quasars only a flux measurement was possible due to the low S/N,
while 15 others are not detected.  For these objects we estimated a
flux upper limit, based on the effective observation length and on the
off-axis distance.  For the 35 remaining objects we extracted and
analyzed the X-ray spectrum.
For all 35 well-detected sources, the background was extracted from a
region of the same observation at the same off-axis distance as the
source.

%\footnote{We
%took into account for this 5\% of sources lying in these areas in 
%our estimates in Table~1}.

\section{First results}

In order to obtain homogeneous results, we fitted all the
0.5-10~keV spectra with a simple model of a power law (of photon index
$\Gamma$) absorbed by an intrinsic column density (N$_H$),
and the Galactic column density from the maps collected by Dickey \&
Lockman (1990). We
used XSPEC V.11.3 (Arnaud 1996) as the fitting engine,
%For spectra with S/N$<$10 we made use of the Cash statistic (Cash
%1979).  For higher S/N spectra, 
and the standard $\chi^2$ minimization technique, after rebinning
in order to have at least 15 counts per spectral channel.  The largest
Galactic column density toward these high Galactic latitude objects is
7$\times$10$^{20}$cm$^{-2}$, and is typically
2-3$\times$10$^{20}$cm$^{-2}$.

The main results of the X-ray spectral and correlation studies are
shown in Figures~1 and 2.  First we note that 
previous
studies were mainly based either on low-redshift objects, like the PG
quasars, or on upper limits on the X-ray flux for objects with $z>1$,
while our sources homogeneously sample the $\sim$0.5-2.5 redshift
range (Fig.~1a), with a detection rate which is high ($\sim70$\%) and
constant with redshift up to $z\sim2.5$ (Fig.~1b). \\

The XMM results are
consistent with the VO3 $\alpha_{OX}$ - luminosity correlation (Fig.~1c),
%(P$_{corr}$=0.XXX), 
and the lack of correlation between $\alpha_{OX}$
and redshift (Fig.~1d). The main new results are the following:
% (P$_{corr}$=0.XXX).

\begin{enumerate}
\item
No spectral evolution with luminosity or redshift has been found.  The
photon index is on average consistent with values $\Gamma=1.8-2$
(Fig.~2a and 2b), typical of low luminosity, low redshift AGNs (George et al.~2000,
Mineo et al.~2000).
Since the rest frame energy range changes substantially with redshift,
reaching above 30~keV for the three z$>$2 quasars, all of which have
well determined $\Gamma$ values, we conclude that quasar spectral
slopes are unchanged up to $\sim$30~keV.

\item
No correlation is present between $\Gamma$ and the B-R color
(Fig.~2c). This implies that the X-ray index and optical colors are
driven by different physical parameters, not by obscuration.  This is
expected e.g. in a disk-corona scenario (e.g. Haardt \& Maraschi~1991), where the
optical-UV slope is determined by the physical conditions in the
accretion disk, while the slope of the X-ray emission is related to
the temperature and Compton depth of the hot corona.

\item
$\Gamma$ is not related to $\alpha_{OX}$ (Fig.~2d).  The implication
is that the {\em efficiency} of X-ray production relative to the
optical (accretion disk) power is separate from the {\em mechanism}
that determines the spectral shape of the X-ray emission.
%
%[OK IN DISK-CORONA MODELS?? EFFICIENCY DEPENDS ON OPTICAL DEPTH.
%BUT SHAPE DEPENDS ON OPTICAL DEPTH AND KT. CAN KT CLEVERLY BALANCE TAU?]
%
Moreover, the spread in $\alpha_{OX}$ cannot be due to absorption, because
then flatter spectra would be found in X-ray weak objects.

\item
The scatter around the $\alpha_{OX}$-optical luminosity relation 
does not depend on optical color. In Fig.~1e we plot the
difference $\Delta\alpha_{0X}$ between the measured $\alpha_{OX}$
and the one expected based on the V03 relation, versus the optical
color B-R.  No correlation is present. If red optical
colors were due to reddening by cold material then red objects would be
X-ray weak due to absorption.  The lack of correlation suggests that
on average the X-ray loudness of quasars and the optical colors are
driven by different physical processes, and are not related to
absorption/extinction.

\item  
All but three quasars show no low-energy photoelectric cut-off
(N$_H<$10$^{22}$cm$^{-2}$, Fig.~1f). Since all these quasars are type~1
objects (i.e. have broad optical emission lines) this is not surprising as
such objects are typically unobscured.  The three exceptions are
the three reddest objects, with (B-R)$>$1 (Fig.~2c) 
and are discussed below.
%is again
%consistent with this Unified Scheme result (see review by Antonucci
%1993).
\end{enumerate}

%%%%%%%%%%%%%%%%%%%%%%%%%%%%%%%%%%%
\subsection{Unusual quasars}

Four of the 35 quasars with XMM X-ray spectra ($\sim10$\% of the
sample) have properties that clearly diverge from the rest. 

Three of them are obscured by a column density
$N_H>10^{22}$~cm$^{-2}$.  Of these, two (SDSS J033718.81+003303.7,
z=0.437, and SDSS J09345834+6112343, z=0.245, marked as \#1 and \#2 in
the figures) are high-z analogs of Narrow Emission Line
Galaxies (Ward et al.~1978) in the local Universe: they have optical
spectra of type 1.8-1.9 Seyferts, an X-ray column density of
$N_H=1.3_{-0.8}^{+0.5}\times10^{22}$~cm$^{-2}$ and
$N_H=2.6_{-0.9}^{+1.8}\times10^{22}$~cm$^{-2}$, respectively, and
moderate 2-10~keV luminosities, $\sim10^{43}$~erg~s$^{-1}$.

The third object (SDSS~J030238.16+000203.4, z=1.35, marked as \#3 in
the figures) is more unusual, in having an optical spectrum typical of a
blue broad line quasar (FWHM(MgII,CIII])$\sim4,000$~km~s$^{-1}$), with an
X-ray spectrum absorbed by an intrinsic column density
$N_H=2.5_{-0.9}^{+1.2}\times 10^{22}$~cm$^{-2}$. The intrinsic
2-10~keV luminosity is $1.5\times10^{44}$~erg~s$^{-1}$.  The absorber
is significantly more dust-free than normal in AGNs, with an upper
limit for the dust to gas ratio $\sim1/50$ of the Galactic value, 
compared to typical AGN values of $\sim$1/10 (Maccacaro, Perola
\& Elvis 1982, Maiolino et al. 2001). These properties make this
object one of the few known high redshift X-ray absorbed quasars.

The fourth interesting object is the only one with a spectral index
significantly flatter than the average (SDSS~J093533.01+612738.6,
$\Gamma=0.6\pm0.2$, Fig.~2, marked as \#4 in the figures).  This
object is classified as a z=0.475 quasar. However, there are no clear
spectral lines in the SDSS spectrum, and the X-ray spectrum, in
addition to a flat continuum, shows a strong emission line
(EW=1.2$\pm0.4$~keV) consistent with an iron K$\alpha$ line at
z$\sim1.4$ (Fig.~3). If confirmed, this would be one of the few high redshift
Compton thick AGNs (e.g. Norman et al.~2002), and a rare case of a
redshift being first determined in X-rays.

%%%%%%%%%%%%%%%%%%%%%%%%%%%%%%%%%%%%%%%%%%%%%%
\subsection{Predicted final SDSS-XMM sample}

We used the results obtained above to estimate the properties
and size of the SDSS-XMM sample once all the SDSS spectra are released,
assuming just the current 5~years of XMM operations. For each of the
sources of the intial sample of 150 objects (Sect.~2) which we did not
analyze directly, we estimated the expected S/N taking into account
the mirror vignetting, the average background counts, the probability
of a non-detection due to a bad position in one or both detectors, and
the probability of background flares affecting part of the
observations.  

In Table~1 we present the statistics obtained for the first 55 objects
(numbers in parentheses), and that expected for the final SDSS-XMM
quasar sample.  We divided the sources in 5 quality groups, according
to the measured (for the 55 objects with X-ray analysis) or expected
(for the other objects) S/N: (1) S/N$>$25: sources for which a
relatively detailed spectral analysis is possible; (2) 10$<$S/N$<$25:
sources for which a basic spectral analysis (power law continuum,
absorption) can be performed; (3) 5$<$S/N$<$10: only a single
component can be fitted (equivalent to a hardness ratio measurement);
(4) 2$<$S/N$<$5: only a flux measurement can be obtained; (5)
non-detection: an upper limit of the flux can be estimated. We also
show the expected distribution in redshift.

%%%%%%%%%%%%
%\begin{table}
%\caption{Sample properties}
%\centerline{\begin{tabular}{lccc|lccc}
%\hline
%S/N$^a$       & S$^b$ & Total$^c$ & Cum.$^d$ & z    & S$^b$ & Total$^c$ & Cum.$^d$  \\
%\hline
%$>$25         &  8  & 59   & 59            & $<0.5$& 6  & 82  & 82 \\  
%10-25         & 23  & 300  & 359 (spectra) & 0.5-1 & 16 & 221 & 303 \\     
%5-10          & 4   & 231  & 590 (colors)  & 1-1.5 & 13 & 179 & 482 \\      
%2-5           & 5   & 64   & 654 (fluxes)  & 1.5-2 & 14 & 192 & 674 \\      
%$<$2          & 15  & 102  & 756           & $>2$  & 6  & 82  & 756 \\      
%\hline
%\end{tabular}}
%
%$^a$: Signal to noise ratio in the 0.5-10 keV band (PN + MOS instruments).
%
%$^b$: Number of sources in the sample presented in
%this paper (55 sources$^6$)
%$^c$: Number of sources expected in the complete sample. 
%$^d$: Cumulative number of sources expected in the complete sample.
%\end{table}
%%%%%%%%%%%
%
%%%%%%%%%%%%%%%%%%%
\begin{table}
%%%%%%%%%%%%%%%%%%%

\caption{SDSS-XMM Sample Distribution}
\centerline{\begin{tabular}{c|c|c|c|c|c|c|c|c|}
%\hline
z$\rightarrow$       & $<0.5$ & 0.5-1 & 1-1.5 & 1.5-2 & $>$2 & $\Sigma_z$ & $\Sigma_z^{cum}$\\
S/N$\downarrow$      &       &       &       &       &      &           &      \\
\hline
$>$25                &(0)5  &(4)26  &(0)5    &(2)10   &(2)15   &  (8)61    & 61  \\
\hline
10-25                &(3)50 &(4)77  &(8)107  &(7)87   &(1)10   &  (23)331  & 392  \\
\hline
5-10                 &(1)5  &(1)26  &(1)51   &(0)87   &(1)57   &  (4)226   & 618  \\
\hline
2-5                  &(0)5  &(2)5   &(1)15   &(1)10   &(1)15   &  (5)50    & 668  \\
\hline
$<$2                 &(2)5  &(4)26  &(4)21   &(4)21   &(1)15   &  (15)88   & 756  \\
\hline
$\Sigma_{S/N}$       &(6)70 &(15)160 &(14)199&(14)215 &(6)112 &  (55)756  &    \\
\hline
$\Sigma_{S/N}^{cum}$ &70    &230     &429    &644     &756    &           &    \\
\hline
\end{tabular}}
\footnotesize{
Number of sources in a given interval of redshift and S/N in
the 0.5-10~keV band (from the combined MOS+PN instruments).
Numbers in parentheses refer to the sample of 55 sources analyzed in this paper.
The other numbers are our predictions for the final SDSS-XMM sample.
The last line contains the cumulative number of quasars up to the
indicated redshift. Similarly, the last column shows the cumulative number of
spectra with a S/N higher than the indicated value.}
%$^a$: Signal to noise ratio in the 0.5-10 keV band (PN + MOS instruments).
%
%$^b$: Number of sources in the sample presented in
%this paper (55 sources$^6$)
%$^c$: Number of sources expected in the complete sample. 
%$^d$: Cumulative number of sources expected in the complete sample.
%
%%%%%%%%%%%%%%%%%%%
\end{table}
%%%%%%%%%%%%%%%%%%%

These numbers will increase as XMM observes longer. A reasonable
estimate is that at the time the full SDSS quasar sample is released
(early 2006) the XMM public archive will be 30-40\% larger than at the
time we performed our search.  The total number of SDSS quasars with
{\em XMM-Newton} detections will then approach 1,000.

\section{Conclusions}

We have cross-correlated the SDSS DR1 quasar sample with the {\em
XMM-Newton} public archive, and found 150 SDSS objects that were
serendipitously observed by XMM. We estimate that over half of these
objects will have a good enough spectrum (S/N$>$10) in order to
estimate the basic spectral parameters (observed 0.5-10~keV photon
index, $\Gamma$, intrinsic photoelectric absorption, N$_H$).
In this paper we presented the analysis of the XMM spectra of 55 of
these quasars observed in long ($>$20~ksec) {\it XMM-Newton}
observations. The quality of the data fully confirms our expectations.

We find that most SDSS quasars have spectral properties analogous to
those of bright PG quasars, with a small spread in $\Gamma$, and
no evolution of $\Gamma$ with redshift ($z$) or luminosity. As there
are three quasars with well-determined $\Gamma$ at $z>$2, we conclude
that quasar spectra have no change in slope up to $\sim$30~keV.
We also find that the X-ray photon index ($\Gamma$) is independent
from X-ray loudness ($\alpha_{OX}$), so that the efficiency of X-ray
production in quasar appears to be independent from the details of the
mechanism of X-ray production.
No correlation is found between $\alpha_{ox}$ and optical colors,
suggesting that the observed spread in the X-ray to optical ratio is
not related to obscuration.

Four of the 35 sources, 10\% of the sample, have unusual X-ray
spectra. The three optically reddest objects in our sample (B-R$>$1)
are also those that show significant intrinsic X-ray absorbing column
density.  Their SDSS spectra show two to be type 1.8-1.9 Seyferts,
while one is an optically normal unreddened quasar, yet with
$N_H\sim$2-3$\times$10$^{22}$~cm$^{-2}$, suggesting the presence of a
dust-free absorber.  One object shows a reflection-dominated X-ray
spectrum, and possibly belongs to the elusive class of high-z,
Compton-thick AGNs.

The work presented here is a first part of a large project. When the
final SDSS quasar sample becomes available, over 700 quasars will have
XMM observations.  An analysis of the recently released DR3 sample
(Abazajian et al.~2005) is on-going, and will include correlations between the X-ray
spectra and the SDSS optical spectral properties.

\acknowledgements 
This work has been partially supported by NASA grant NAG5-16932.

%%%%%%%%%%%%%%%%

%%%%%%%%%%%%%%%

\clearpage

%%%%%%%%%%%%%%%
\begin{figure}
\epsscale{0.9}
\plotone{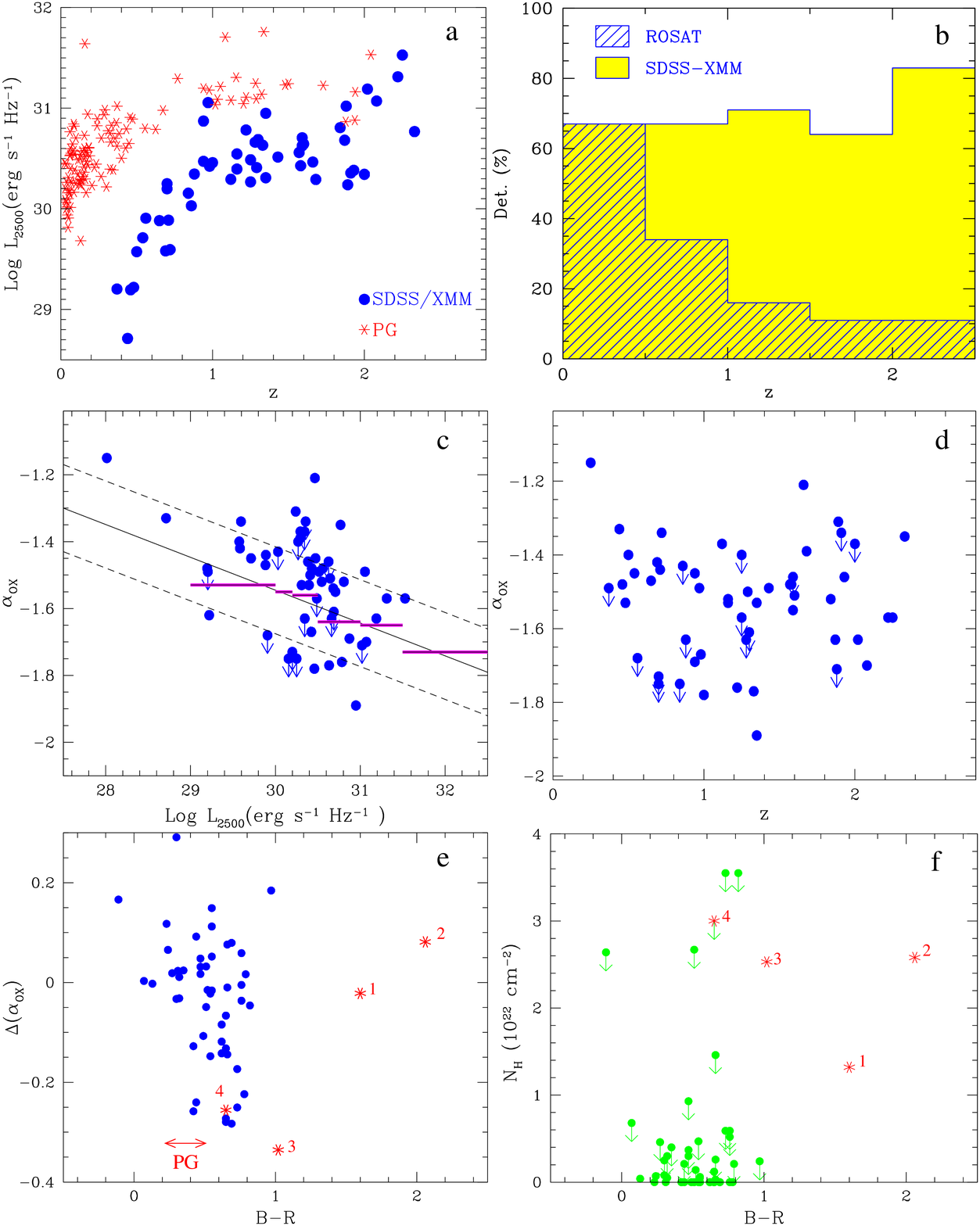} 
\figcaption{\footnotesize{Results of X-ray analysis and correlation
with optical properties for the SDSS-XMM sample. 
{\bf a}: Luminosity-redshift plot for our sample, compared with
that of PG quasars.
{\bf b}: Fraction of detected
sources versus redshift for our sample and the one of Yuan et al. (1998).
{\bf c} and {\bf d}: 
X-ray photon index versus optical 2500~\AA~monochromatic
luminosity and versus redshift. In the first plot, our data are
compared with the correlation inferred by V03 (continuous
thick line, with the two dashed lines indicating the statistical errors,
and with the correlation by Yuan et al. 1998 (horizontal segments).
{\bf e}: 
%and {\bf f}:
Residuals of the measured $\alpha_{OX}$  with respect to the V03 relation,
versus optica B-R color.
{\bf f}: $N_H$ versus B-R color. The four peculiar objects discussed in Section~3.1
are marked as stars.
 }}
\end{figure}
%%%%%%%%%%%%%%%

\clearpage

%%%%%%%%%%%%%%%
\begin{figure}
\plotone{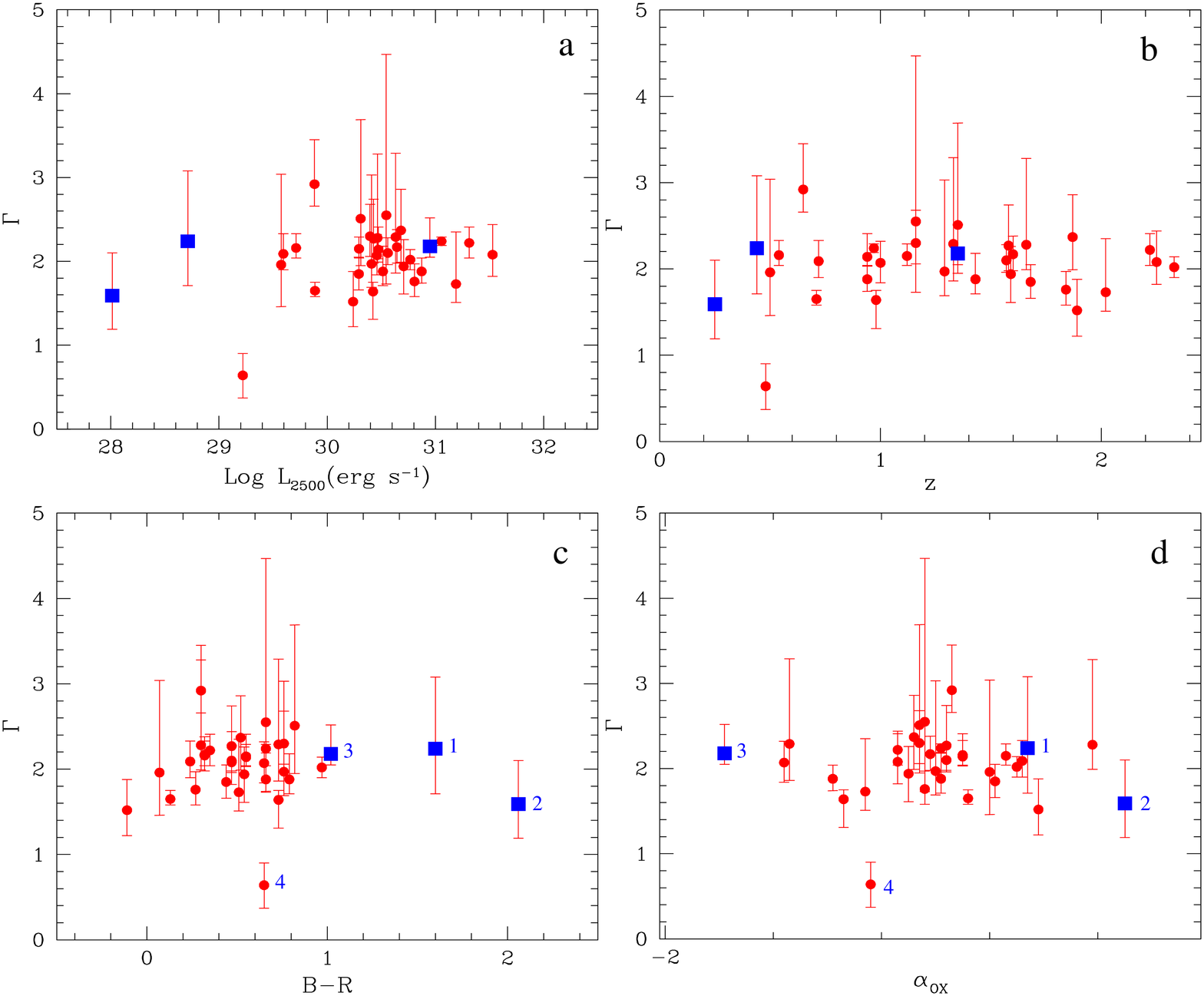} 
\figcaption{X-ray photon index versus {\bf a}:
optical 2500~\AA~monochromatic
luminosity; {\bf b}: redshift; {\bf c}: B-R color; {\bf d}: $\alpha_{OX}$.
The three squares are for the ``peculiar'' quasars 
with a measured column density $N_H>10^{22}$~cm$^{-2}$.}  
\end{figure}
%%%%%%%%%%%%%%%

\begin{figure}
\includegraphics[angle=-90,width=16cm]{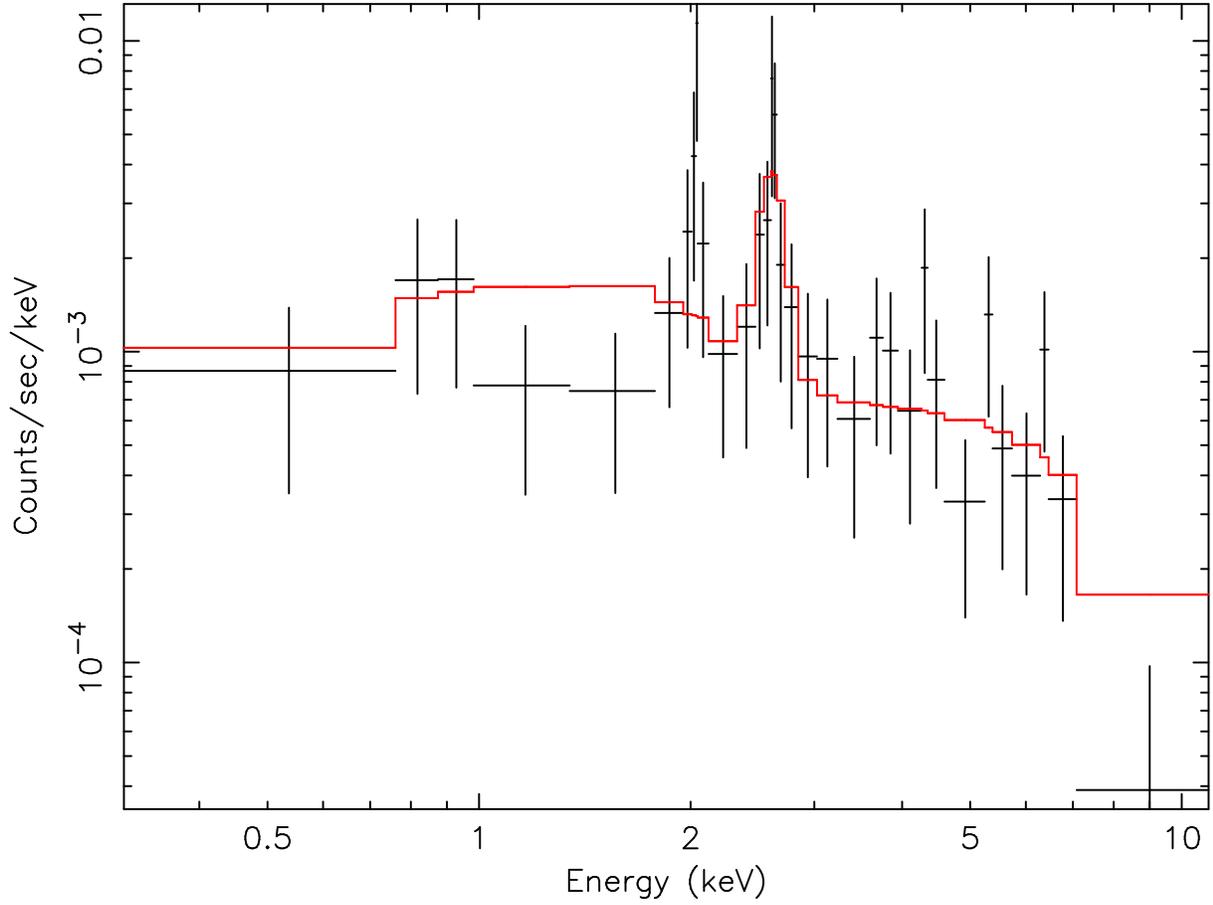} 
\figcaption{X-ray spectrum of the object marked as \#4. Both the flat
continuum ($\Gamma\sim0.7$) and the putative iron line (4$\sigma$ excess, 
with EW$\sim1.2$~keV) point to a reflection dominated, $z\sim1.4$ quasar}.
\end{figure}

%%%%%%%%%%%%%%%
\end{document}